\documentclass{vgtc}                          

\ifpdf
  \pdfoutput=1\relax                   
  \usepackage{graphicx}                
  \DeclareGraphicsExtensions{.pdf,.png,.jpg,.jpeg} 
\else
  \usepackage{graphicx}                
  \DeclareGraphicsExtensions{.eps}     
\fi%

\graphicspath{{figures/}{pictures/}{images/}{./}} 

\usepackage{microtype}                 
\PassOptionsToPackage{warn}{textcomp}  
\usepackage{textcomp}                  
\usepackage{mathptmx}                  
\usepackage{times}                     
\usepackage{cite}                      
\usepackage{tabu}                      
\usepackage{booktabs}                  
\usepackage{multirow}
\usepackage[dvipsnames, table,xcdraw, svgnames]{xcolor}

\usepackage{enumitem}

\onlineid{5350}

\vgtccategory{Research}

\vgtcinsertpkg

\title{Edge-Centric Space Rescaling with Redirected Walking \\ for Dissimilar Physical-Virtual Space Registration}

\author{Dooyoung Kim\thanks{e-mail: dooyoung.kim@kaist.ac.kr}\\ %
         \scriptsize KAIST UVR Lab. %
 \and Woontack Woo\thanks{Corresponding Author. e-mail: wwoo@kaist.ac.kr}\\ %
      \parbox{1.4in}{\scriptsize \centering KAIST UVR Lab. \\ KAIST KI-ITC ARRC}}

\teaser{
  \centering
  \includegraphics[width=\linewidth]{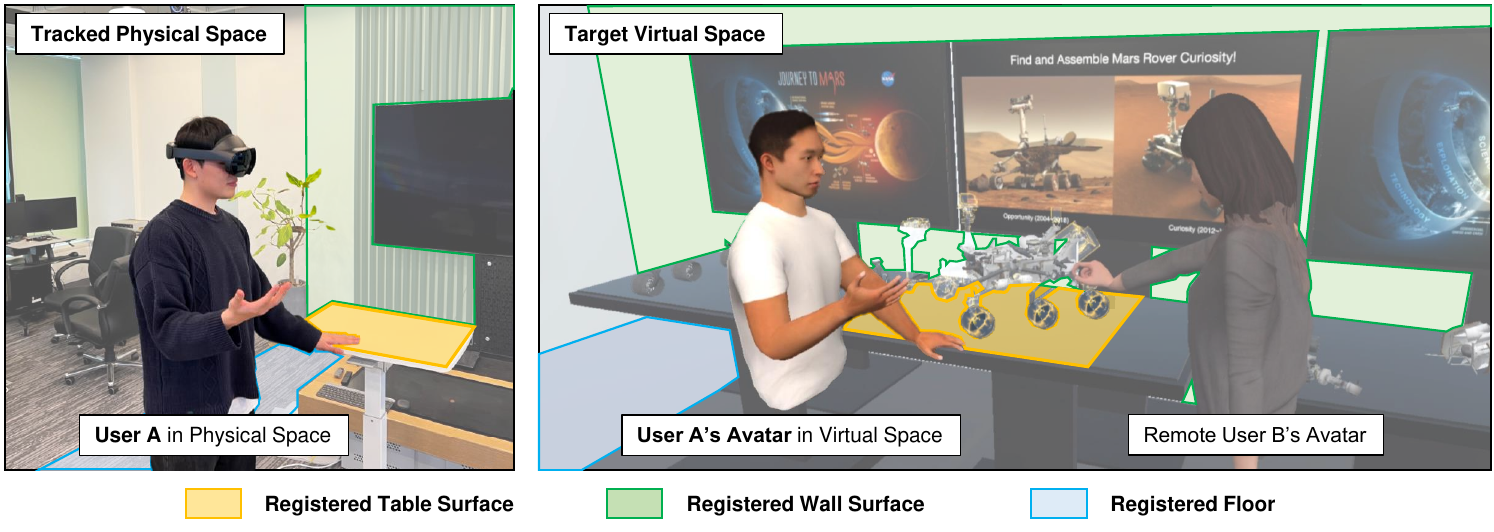}
  \caption{Our space-rescaling technique aligns edges and planes between physical and virtual space precisely and increases the registered areas, so the user can physically interact with objects and planes in virtual space. User A (left) connected to the target virtual space (right) dissimilar from his physical space through his avatar. He could physically interact with the registered floor, table, and wall. Each colored area refers registered area: Orange - table (main object) surface, green - wall surface, blue - movable floor.}
  \label{fig:teaser}
}

\abstract{
We propose a novel space-rescaling technique for registering dissimilar physical-virtual spaces by utilizing the effects of adjusting physical space with redirected walking. Achieving a seamless immersive Virtual Reality (VR) experience requires overcoming the spatial heterogeneities between the physical and virtual spaces and accurately aligning the VR environment with the user's tracked physical space. However, existing space-matching algorithms that rely on one-to-one scale mapping are inadequate when dealing with highly dissimilar physical and virtual spaces, and redirected walking controllers could not utilize basic geometric information from physical space in the virtual space due to coordinate distortion. To address these issues, we apply relative translation gains to partitioned space grids based on the main interactable object's edge, which enables space-adaptive modification effects of physical space without coordinate distortion. Our evaluation results demonstrate the effectiveness of our algorithm in aligning the main object's edge, surface, and wall, as well as securing the largest registered area compared to alternative methods under all conditions. These findings can be used to create an immersive play area for VR content where users can receive passive feedback from the plane and edge in their physical environment.

} 


\CCScatlist{
  \CCScatTwelve{Human-centered computing}{Human Computer Interaction (HCI)}{Interaction paradigms}{Virtual reality};
}

\begin{document}

\firstsection{Introduction}

\maketitle

Spatial computing devices have revolutionized the manner in which users engage with virtual environments, offering an unprecedented level of realism and immersion~\cite{orts2016holoportation, skarbez2021virtual}. Augmented Reality (AR), Virtual Reality (VR), and Mixed Reality (MR) Head Mounted Displays (HMDs) extend beyond the confines of traditional 2D screens and video communications, fostering a sensation of shared space coexistence~\cite{piumsomboon2017exploring, schafer2022survey}. For users to truly immerse themselves in these virtual environments, the synchronization between the virtual and the user's physical space, considering spatial disparities, becomes pivotal. Notably, many current VR HMDs predominantly utilize predefined void spaces as interactive zones, constraining mobility and efficient space utilization\footnote{https://www.meta.com/help/quest/articles/in-vr-experiences/oculus-features/oculus-guardian/}. This modus operandi often presupposes a stationary user, thereby stifling the evolution of immersive VR narratives. Addressing these constraints necessitates a space registration mechanism that capitalizes on the physical space's geometric details and can adapt to shifts in user postures.

Registering virtual with physical spaces presents a significant challenge, particularly when addressing spatial disparities and guaranteeing ample space for user movement and interaction. Traditional avatar positioning methods relying on forced mapping have been known to cause communication discrepancies stemming from avatar misalignment~\cite{yoon2021full, congdon2018merging, tomar2019conformal, yoon2020placement}. Additionally, redirected walking controllers occur coordinate warping, constraining the effective incorporation of real-world geometric data in virtual settings~\cite{lee2019real, chang2021redirection, williams2021arc}. Although alignment-centric techniques offer the potential to synchronize coordinate systems, they encounter challenges in providing adequate space, especially when dealing with markedly differing spaces or increasing space count~\cite{keshavarzi2020optimization, keshavarzi2022mutual, lehment2014creating}. Consequently, a robust space registration approach should facilitate spatial modifications while preserving a consistent coordinate framework.

This research introduces a method that adjusts physical space dimensions using relative translation gains (RTGs)~\cite{kim2021adjusting} derived from redirected walking. As depicted in \autoref{fig:teaser}, users in VR can interact with their physical surroundings in a synchronized virtual realm. For the space-adaptive rescaling effect, the RTGs were applied to each space grid within threshold range~\cite{kim2022configuration}, and redirection gains between grids were linearly smoothed. Moreover, new spatial registration metrics were proposed to quantitatively measure the alignment of edges and planes between virtual and physical spaces, as physical touch interaction is beneficial for an immersive VR experience~\cite{hoffman1998physically, cheok2002touch}. An optimization algorithm with an objective function generated with four spatial registration metrics can achieve optimal spatial registration. Furthermore, we formulated spatial complexity (SC($E$)) that can reflect not only object density but also space size and object distribution. By comparing SC($E$) from physical and virtual spaces, spatial dissimilarity (SD($E_{virt}$, $E_{phys}$)) and spatial matching difficulty (SMD($E_{virt}$, $E_{phys}$)) were presented to help design structured experiment by leverage the difficulty level of registering the target virtual space to the tracked physical space.

We conducted an evaluation using four space pairs for XR remote collaboration scenarios and validated our proposed method (RTG-Grid) against two other spatial registration methods (RTG-Single, 1:1 Scale). RTG-Single refers to applying a single RTG to the entire space, and 1:1 Scale refers to spatial matching with the original space scale. The data underscored RTG-Grid's superiority in aligning the target collaboration object's edges and surfaces and maximizing the registered space in every tested scenario. Moreover, we validated the efficacy of SC($E$), SD($E_{virt}$, $E_{phys}$), and SMD($E_{virt}$, $E_{phys}$) as tools to gauge spatial details and the challenge of aligning physical and virtual realms, juxtaposing them against previously introduced metrics for spatial complexity from ARC and ENI~\cite{williams2021arc, williams2022eni}. From our analyses, we distilled three insights for developers contemplating optimization algorithms for aligning heterogeneous virtual and physical environments: First, RDW can serve as a tool to rescale spaces, ensuring precise alignment of boundaries and surfaces. Second, the objective function's primary factor should be set according to the main interaction object. Lastly, it is important to focus on registering the subspace where users will interact most rather than aiming to map the entire space.

The contribution of our study can be categorized into three main areas. Firstly, we introduced a novel edge-centric space-rescaling technique with length-based RDW for registering virtual space to the user's tracked physical space. Contrary to prior techniques, we emphasize aligning primary interaction points and their respective subspaces, rather than solely amplifying overall semantic alignment. Secondly, to enable spatial matching according to VR scenarios, we proposed four quantitative metrics for spatial registration to analyze the spatial matching results of dissimilar physical-virtual spaces. Lastly, we developed equations to quantitatively measure the spatial complexity, spatial dissimilarity, and spatial matching difficulty to assess the level of registering dissimilar physical-virtual spaces. By leveraging the optimization algorithm and evaluation metrics proposed in this study, developers can achieve optimal spatial registration by adjusting the weights of optimization terms based on their VR content goals using our objective function and implications.

\section{Background}

\subsection{Space Rescaling with Redirected Walking}

The redirected walking (RDW) technique in VR aims to increase the user's movable space by adjusting their visually perceived locomotion within certain thresholds~\cite{razzaque2005redirected, rietzler2018rethinking, steinicke2009estimation, nguyen2020effect, grechkin2016revisiting, williams2019estimation}. Various RDW controllers have been developed to maximize the navigable area while minimizing collisions and resets~\cite{bachmann2019multi, chang2021redirection, dong2020dynamic, lee2019real, strauss2020steering, xu2023multi}. The simple controllers try to steer the user to a certain target~\cite{messinger2019effects, lee2020optimal}, and there were alignment-based RDW controllers that consider physical and virtual space compatibility, enabling more space-adaptive redirection~\cite{williams2021redirected, williams2021arc}. However, existing controllers focus on reducing collisions and resets, leading to abrupt changes in gain values and uncomfortable experiences~\cite{sakono2021redirected}. To address this, we propose creating a redirection map that ensures smooth gain transitions while utilizing the rescaling effects of RDW to expand the available space.

Although most controllers commonly employ rotation and curvature gains due to their significant impact on altering spatial perception, rotational gains introduce challenges in aligning real-world edges and flat surfaces with the virtual environment due to coordinate discrepancies. Since enabling physical object interaction by aligning the edge and plane between physical and virtual spaces could achieve the immersive VR experience~\cite{hoffman1998physically, cheok2002touch}, translation gain is appropriate for using RDW as a space modification. To do so, relative translation gains (RTGs), which apply different translation gains to the width and depth in space, can be used for space-adaptive space rescaling~\cite{kim2021adjusting}. Although the concept of matching heterogeneous spaces with RTG was presented before~\cite{kim2022mutual}, further research is needed to investigate a more space-adaptive rescaling method considering the spatial configuration and how to apply this space-rescaling technique to registering dissimilar space.

\subsection{Dissimilar Space Mapping and Registration}

The registration of virtual spaces to physical spaces poses a significant challenge in VR experiences, especially when the configurations of the two spaces differ significantly. Previous avatar position mapping approaches aimed to connect every position between the two spaces, but this may not be feasible when the spaces are highly dissimilar~\cite{yoon2021full, congdon2018merging, herskovitz2022xspace, tomar2019conformal}. Furthermore, avatar retargeting approaches have attempted to correlate user movement from his/her physical space to the target virtual space, but they resulted in unexpected shifts and compromised immersion~\cite{yoon2020placement}. On the other hand, Wang et al.~\cite{wang2022predict} attempted to match user movement smoothly between the tracked and target spaces by correlating objects of interest that the user wishes to move, which could result in inaccurate user pointing and gazing direction due to coordinate distortion. Therefore, one of the important objectives of space registration should be to generate a continuous repositioning map while aligning the coordinate system.

Object-centric space-matching methods such as projecting a remote user onto their physical sofa~\cite{pejsa2016room2room} or retargeting a round table and square table by grid-based mapping~\cite{de2022adaptables} have also been developed. The VR-oriented spatial matching algorithm was also proposed, which could automatically reconstruct the target virtual space in the user's smaller and different physical space~\cite{dong2021tailored}. Alignment-based methods, such as space-aligning and object-centric approaches, have been proposed to maintain the coordinate system between spaces~\cite{lehment2014creating, keshavarzi2020optimization, keshavarzi2022mutual}. These methods could support users in different states by aligning semantic information, but they may result in a narrow shared space or limited adaptability to dissimilar configurations. On the other hand, these alignment-based spatial matching methods primarily consider semantic match ratio or registered space size~\cite{lehment2014creating, keshavarzi2020optimization}, but spatial registration results should be validated from various perspectives considering the context of the VR experience. To this end, quantitative spatial registration metrics that allow developers to change spatial alignment's target goals according to the user scenario should be developed.

\subsection{Environmental Complexity and Dissimilarity}

Registering dissimilar virtual space to a target physical space is a challenging task that requires understanding the spatial heterogeneity between the two spaces. Quantifying spatial dissimilarity is difficult due to its qualitative nature, such as relationships between objects and spatial layout. However, certain quantifiable aspects like space size, object density, and object distribution can provide essential information for spatial matching. Previous methods focused on measuring openness or environmental complexity but had limitations in comparing spatial configurations or considering object densities~\cite{fisher2003spatial}. Additionally, environmental complexity proposed in the robotics community focuses on distinguishing the difficulty of motion finding, but mostly considered empty navigable areas~\cite{anderson2007proposed, shell2003human}.

In AR and VR research, spatial complexity about the user's experience has been studied, but without quantitative measurement~\cite{shin2021user, shin2022effects}. Existing spatial comparison metrics from ENI~\cite{williams2022eni} could measure the navigability of dissimilar virtual space from the user's physical space, but it's hard to reflect spatial dissimilarity. The environmental complexity (CR) was proposed in ARC~\cite{williams2021arc} but was mostly influenced by object density. To this end, the spatial complexity term, which considers not only space size but also space size and object dispersion, is needed. Through quantitative comparison between dissimilar spaces, researchers could design a more structured study and validate spatial registration results more constructively.

\section{Methodology}

This study proposes an edge-centric space-rescaling technique for registering virtual space to physical space by utilizing basic geometric information such as floors, walls, and horizontal surfaces. We begin by presenting a way to make an effect of rescaling physical space with relative translation gains (RTGs) in RDW. The physical space was divided into grid units for space-adaptive rescaling, and different pairs of RTGs were applied to each grid. Then, we presented quantitative evaluation metrics for space registration to validate how well virtual space is registered to the physical space. In addition, we proposed a spatial complexity (SC($E$)) to measure the environment's complexity considering space size, object density, and object distributions. With SC($E$), we formulated spatial dissimilarity (SD($E_{virt}, E_{phys}$)) and spatial matching difficulty (SMD($E_{virt}, E_{phys}$)) to measure the difficulty level of registering target virtual space to the tracked physical space.

\subsection{Space Rescaling with Relative Translation Gains}

\begin{figure}[t]
 \centering
 \includegraphics[width=\columnwidth]{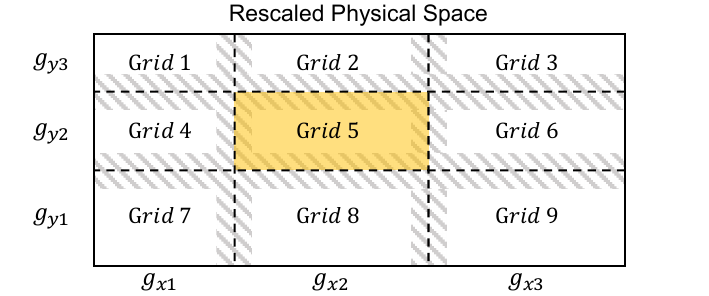}
 \caption{Top-view of the physical space for rescaling. Grids were divided by the extension of the orange-colored main object (Grid 5)'s four edges. Dashed areas indicate the linear gain smoothing area.}
 \label{fig:Grid}
\end{figure}

When we interpret the RDW in terms of space, it can be explained as the effects of rescaling the VR user's physical space. To modify physical space while maintaining a unified coordinate system between the two spaces, we utilized RTGs in RDW~\cite{kim2021adjusting}. RTG is a translation-gain-based RDW that allows space-adaptive rescaling by applying different gains to the width and depth of the space. As the RTG threshold is influenced by spatial configuration, we employed the findings of Steiniche et al.~\cite{steinicke2009estimation} and Kim et al.~\cite{kim2022effects,kim2022configuration} together to set the threshold range considering the target virtual space's spatial configuration. RTG consists of an x-axis translation gain ($g_{x}$) and a y-axis translation gain ($g_{y}$). The RTG threshold is defined with a 2D translation ratio ($\alpha_{T}$), which is the ratio of $g_{x}$ to $g_{y}$, and the physical space's scale could change within $\alpha_{T}$.

Since it is challenging to match every object in dissimilar virtual-physical environments, we concentrated on matching the particular object, which is the main target of interaction with the given VR scenario. It can be extended to multiple target objects, but for the sake of clarity, we explained a method that divides the space based on a single object and adaptively transforms it. The object of interest was named as ``main object", and the physical space was partitioned into nine grids centered on this main object's edge. As depicted in \autoref{fig:Grid}, the edges of the main object were extended to create a total of nine grids, and nine RTG pairs generated from the combination of three x-axis translation gains ($G_{x} = [g_{x1}, g_{x2}, g_{x3}]$) and three y-axis translation gains ($G_{y} = [g_{y1}, g_{y2}, g_{y3}]$) were applied to each grid. By ensuring the maximum difference between $G_{x}$ and $G_{y}$ falls within the RTG threshold range, the other RTG pairs will also be within the corresponding threshold range. Therefore, the inequality equations for the RTG threshold can be written as follow:

\begin{equation}
    0.86 < g_{x},\,g_{y} < 1.26.
    \label{equ:thres_basic}
\end{equation}

\begin{equation}
    \ \alpha_{T, l} \leq \frac{min(G_{x})}{max(G_{y})}, \frac{max(G_{x})}{min(G_{y})} \leq  \alpha_{T, h}. \\
    \label{equ:thres_RTG1}
\end{equation}

\noindent Each $g_{x}$ and $g_{y}$ should have a value between 0.86 and 1.26~\cite{steinicke2009estimation}, and the lower boundary $\alpha_{T, l}$ and upper boundary $\alpha_{T, h}$ can be set according to the target virtual space's configuration~\cite{kim2022configuration}. After applying the optimal RTG values for space rescaling, we applied linear gain smoothing for a distance of $2l_s$ m ($l_s$ m before and after passing the grid boundary) to reduce the user's discomfort~\cite{sakono2021redirected}. For example, if a user walks from Grid 1 to Grid 2 (\autoref{fig:Grid}) and is at a distance of $d$ m from the boundary line between the two grids ($-l_s$ m $<$ $d$ $<$ $l_s$ m), the x-axis translation $g_{x}$ will be $0.5((g_{x2} + g_{x1}) + (g_{x2} - g_{x1})d/l_{s})$. \autoref{fig:Grid}'s dashed area indicates the smooth gain change area where RTG linearly changes from one grid to the adjacent grid. 


\begin{figure*}[htb!]
 \centering
 \includegraphics[width=\textwidth]{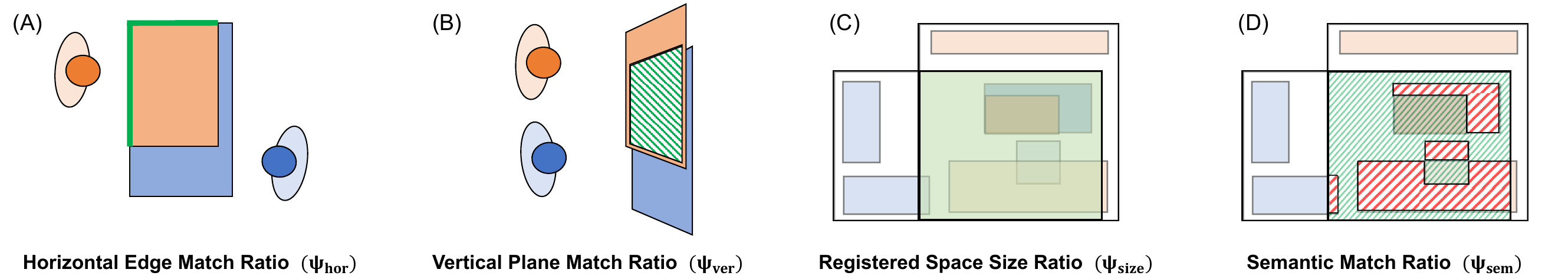}
 \caption{Illustrations for four evaluation metrics for spatial registration. Green lines and dashed areas refer to matched edges, planes, and areas. Red dashed areas refer to semantically unmatched areas. Each metric can be computed with described matched areas or lengths.}
 \label{fig:eval-metrics}
\end{figure*}

\subsection{Evaluation Metrics for Space Registration}

We proposed four essential space-registering evaluation metrics to validate the alignment between virtual and physical spaces as illustrated in \autoref{fig:eval-metrics}. In addition to the two previously used spatial matching metrics (registered space size and semantic match ratio)~\cite{lehment2014creating, keshavarzi2020optimization}, we added two new metrics about surface-centric terms based on previous qualitative research about the effectiveness of surface-centric spatial matching and physical touching in VR~\cite{gronbaek2023partially, zenner2021combining, franzluebbers2018performance}.

\paragraph{Horizontal Edge Match Ratio} The first metric is the horizontal edge match ratio between the virtual and physical objects' horizontal surfaces. For the horizontal plane, the correct alignment of the edges is important from a real-world interaction perspective, so the edge match ratio is set as a key metric rather than a simple plane match ratio. \autoref{fig:eval-metrics}(A) shows the concept of horizontal edge match ratio $\psi_{hor}$ and it can be defined as:

\begin{equation}
    \ \psi_{hor}(G, \phi) = \frac{IL(O^{*}_{virt}, O^{*}_{phys})}{L(O^{*}_{virt})} \\
    \label{equ:hor}
\end{equation}

\noindent where $G$ represents the nine RTG pairs used for rescaling the physical environment, while $\phi$ represents the relative position vector of the virtual environment to the physical environment. $O^{*}_{virt}$ and $O^{*}_{phys}$ are the top-viewed 2D polygon of the target object with a horizontal plane in the virtual and physical environments, respectively. The variable $IL(O^{*}_{virt}, O^{*}_{phys}$) denotes the sum of the intersected edge length between $O^{*}_{virt}$ and $O^{*}_{phys}$, while $L(O^{*}_{virt}$) refers to the sum of the edge length of $O^{*}_{virt}$. $\psi_{hor}$ allows us to measure how much of the edge from the target object in the virtual space can be registered to the edge of the physical object's horizontal plane.

\paragraph{Vertical Plane Match Ratio} The next evaluation metric is the vertical plane match ratio. Vertical planes, such as walls or whiteboards, are among the most common interaction planes, along with horizontal surfaces. If the vertical plane is aligned, users can directly touch the physical wall when they attach virtual content to the wall or whiteboard in the virtual environment. \autoref{fig:eval-metrics}(B) shows the concept of vertical plane match ratio $\psi_{ver}$ and it can be measured with following formula:

\begin{equation}
    \ \psi_{ver}(G, \phi) = \frac{IL(P_{virt}, P_{phys})}{L(P_{virt})} \\
    \label{equ:ver}
\end{equation}

\noindent where $P_{virt}$ and $P_{phys}$ represent the top-viewed polygon from the object (or floor) with the vertical plane in a virtual and a physical environment. Similar to the $\psi_{hor}$, the sum of the mutual intersected top-viewed plane's length between $P_{virt}$ and $P_{phys}$ is indicated by $IL(P_{virt}, P_{phys})$, while the sum of $P_{virt}$'s edge length is $L(P_{virt})$. 

\paragraph{Registered Space Size Ratio} The third metric is the registered space size ratio between the virtual and physical environments. Ensuring a sufficiently large registered area is crucial for users to explore and interact more freely. Figure~\ref{fig:eval-metrics}(C) shows the registered space size ratio $\psi_{size}(G, \phi)$ and it can be defined as follows:

\begin{equation}
    \ \psi_{size}(G, \phi) = \frac{A(I(E_{virt}, E_{phys}))}{A(E_{virt})} \\
    \label{equ:size}
\end{equation}

\noindent where $I(E_{virt}$, $E_{phys}$) represents the intersected polygon resulting from aligning the multi-polygon input of $E_{virt}$ with that of $E_{phys}$. Through $A(I(E_{virt}, E_{phys}))$, the registered space size can be computed, and the $\psi_{size}(G, \phi)$ can be obtained by dividing $A(I(E_{virt}, E_{phys}))$ with the virtual environment's total area, $A(E_{virt})$.

\paragraph{Semantic Match Ratio} The last metric is the semantic match ratio between two spaces. This is the most frequently used metric to measure the performance of spatial matching in previous alignment-based optimization algorithms~\cite{keshavarzi2020optimization, keshavarzi2022mutual, kim2022mutual}, which means the ratio of the semantically aligned area out of the total registered space. Figure~\ref{fig:eval-metrics}(D) shows the semantic match ratio $\psi_{sem}(G, \phi)$ and it can be formalized as follow:

\begin{equation}
    \ \psi_{sem}(G, \phi) = \frac{\sum_{l\in L} IA_l(E_{virt}, E_{phys})}{A(I(E_{virt}, E_{phys}))} \\
    \label{equ:semantic}
\end{equation}

\noindent where $L$ is the set of possible semantic labels and $IA_l(E_{virt}, E_{phys})$ is the area of the intersected region of semantic label $l$ after registering $E_{virt}$ to the $E_{phys}$. The numerator in \autoref{equ:semantic} represents the total area of the intersected regions of all semantic labels, and the denominator represents the total area of the registered space. The resulting ratio $\psi_{sem}(G, \phi)$ ranges from zero to one, where a higher value indicates a better semantic match between $E_{virt}$ and $E_{phys}$.

\paragraph{Objective Function} With the above four metrics, we generated the objective function for the optimization algorithm:

\begin{equation}
    \ O(G, \phi) = \sum_{i=1}^{n}\omega_{1, i}\psi_{hor, i}+\sum_{j=1}^{m}\omega_{2, j}\psi_{ver, j} + \omega_{3}\psi_{size} + \omega_{4}\psi_{sem}. \\
    \label{equ:objFunc}
\end{equation}

\noindent where n and m are the number of horizontal-/vertical-planes to be matched, respectively. Each spatial registration metric is a unitless value between zero and one, with increases in each metric indicating better spatial matching performance in each aspect. By assigning weights to the coefficients $\omega$ in the objective function depending on the intended purpose of registering the virtual environment, the optimal values for physical space scaling ($G$) and position vector ($\phi$) that maximize the objective function can be obtained for registering the target virtual space to the user's physical space.

\begin{figure*}[ht]
 \centering
 \includegraphics[width=\textwidth]{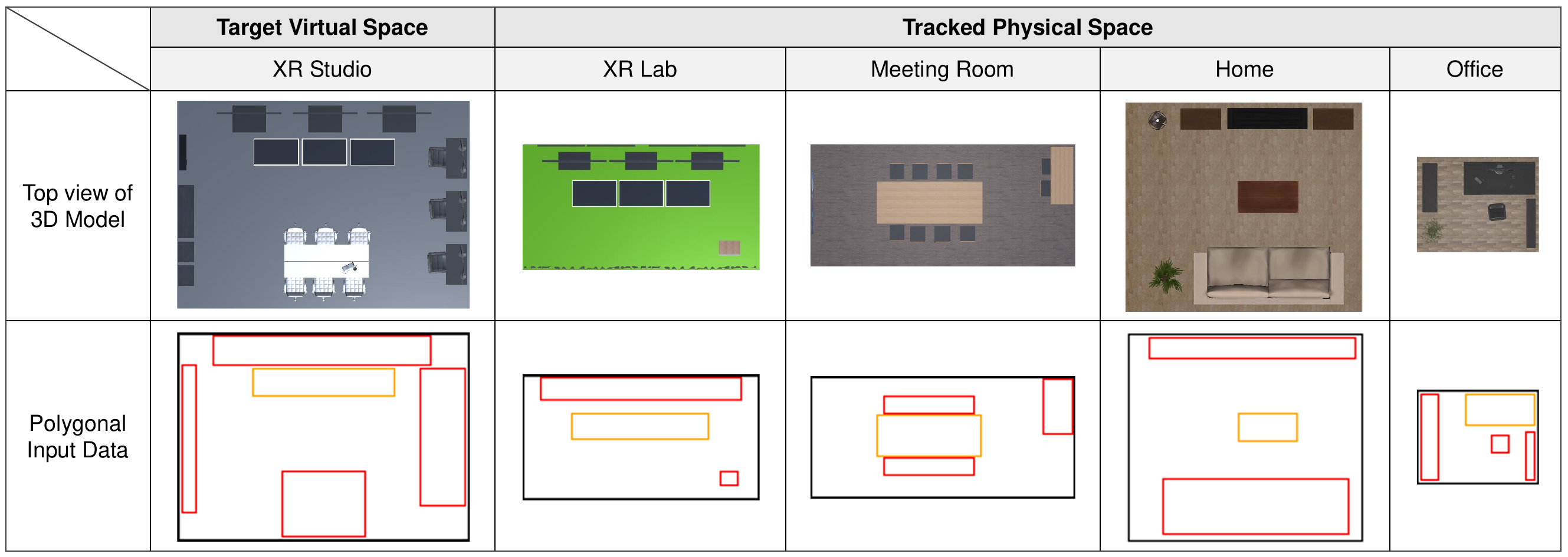}
 \caption{Five input spaces for evaluation. We generated polygonal input data from the top view of 3D indoor space models. The polygon's color refers to the following semantic labeled polygon: Black - floor, orange - the main object (table), red - obstacles.}
 \label{fig:dataset}
\end{figure*}

\subsection{Spatial Complexity and Spatial Dissimilarity}

\paragraph{Spatial Complexity} Since the performance of space registration is highly dependent on the dissimilarity between two spaces, we proposed a way to measure spatial complexity, SC($E$). Drawing on the concept of C($E$)~\cite{williams2021arc}, we aim to construct an SC($E$) metric that increases when the spatial complexity decreases. While C($E$) mainly reflects the density of objects in a room, SC($E$) aims to contain not only object density but also space size and object distribution. To express SC($E$), we decompose it into an area-related term and a spatial layout-related term. Initially, we incorporated the space size component into C($E$) by multiplying the space area since C($E$) already contains some information about space size. However, as SC($E$) is overly sensitive to space size when we directly multiply the environment area, we multiply the square root of the environment area by C($E$) to reduce this effect.

In order to reflect the spatial layout of a virtual environment, we proposed an object scatteredness, OS($E$), which considers the distribution of objects in the space. As the spatial complexity of the environment decreases when objects are more scattered~\cite{shin2021user, shin2022effects}, OS($E$) should increase accordingly to contain the spatial complexity in SC($E$). As the area occupied by objects is already reflected in C($E$), we defined OS($E$) as follows:

\begin{equation}
    \ OS(E) = \left (\frac{\sum_{O_{i}, O_{j} \in O}^{} d(o_{i}, o_{j})}{2N(O)} \right )^2
    \label{equ:OS}
\end{equation}

\noindent where $O$ is the every object in the $E$, and $o_{i}$, $o_{j}$ denote the center positions of the bounding boxes of object $O_{i}$ and $O_{j}$ respectively. $d(o_{i}, o_{j})$ represents the distance between $o_{i}$ and $o_{j}$. The number of objects in $E$ is denoted by N($O$). Since $d(o_{n}, o_{m})$ and $d(o_{m}, o_{n})$ are added twice in the numerator equation even though they refer to the same distance between $O_{n}$ and $O_{m}$, ${\sum_{o_{i}, o_{j} \in O}^{} d(o_{i}, o_{j})}$ is divided with 2N($O$) for the average distance between two object pairs. The unit of OS($E$) is $m^2$, which is consistent with the unit of $ \sqrt{A(E)} \times C$. As demonstrated above, \autoref{equ:OS} does not incorporate object area information but includes the distance between them.

In conclusion, the spatial complexity SC($E$) can be computed with the following equation:

\begin{equation}
    \ SC(E) = \sqrt{A(E)} \times C(E) + OS(E).\
    \label{equ:SC}
\end{equation}

\noindent Here, A($E$) represents the $E$'s total floor area. C($E$) is the complexity term defined in ARC~\cite{williams2021arc}, while OS($E$) represents the object scatteredness. Through \autoref{equ:SC}, we could obtain spatial complexity, which contains environmental information such as space size, object density, and the distribution of objects.

\paragraph{Spatial Dissimilarity} To measure the spatial dissimilarity between physical and virtual environments, we compare their spatial complexity values, denoted by SC($E_{virt}$) and SC($E_{phys}$), respectively. We propose the following formula to calculate the spatial dissimilarity SD($E_{virt}, E_{phys}$) between virtual and physical environment:

\begin{equation}
    \ SD(E_{virt}, E_{phys}) = \left|ln\left (\frac{SC(E_{virt})} {SC(E_{phys})}\right ) \right|. \
    \label{equ:dissimilar}
\end{equation}

\noindent By applying the logistic and absolute functions to the ratio of SC($E_{virt}$) and SC($E_{phys}$), the resulting value is independent of the order of the numerator and denominator. An increase in SD($E_{virt}, E_{phys}$) signifies a higher dissimilarity between the physical and virtual environments.

\paragraph{Spatial Matching Difficulty} In order to measure the difficulty level of aligning virtual and physical spaces that are centered on the target object, we introduced a spatial matching difficulty indicator called SMD($E_{virt}$, $E_{phys}$) based on the SD($E_{virt}$, $E_{phys}$). Since the surface is mostly used as an interaction area, we utilize the main object's surface area between physical and virtual space, and it can be formulated as follows:

\begin{equation}
    \ SMD(E_{virt}, E_{phys}) = \left|ln\left (SD(E_{virt}, E_{phys}) \times \frac{A(O^{*}_{virt})} {A(O^{*}_{phys})}\right ) \right|. \
    \label{equ:SMD}
\end{equation}

\noindent $SMD(E_{virt}, E_{phys})$ is expressed by multiplying the main object's surface area ratio between virtual and physical space to the SD($E_{virt}, E_{phys}$). Similar to the SD($E_{virt}, E_{phys}$), the logistic and absolute functions were applied. Using the SMD from \autoref{equ:SMD}, the difficulty level in aligning virtual and physical spaces centered on the main object can be measured. Similar to SD, an increase in SMD indicates a higher difficulty in registering $E_{virt}$ to $E_{phys}$.

\section{Evaluation}

Our evaluation aims to compare the performance of our proposed spatial matching algorithm focus on a realistic scenario with real spaces as other spatial matching researches evaluated~\cite{keshavarzi2020optimization, yoon2020placement}. Our target scenario was registering a host's virtual meeting room to each client's space for XR remote collaboration. We assumed a table and walls are utilized as the main interactable surface, so our objective function consists of one $\psi_{hor}$ and one $\psi_{ver}$ for each. Additionally, SD and SMD were validated by comparison with metrics about spatial dissimilarity (CR)~\cite{williams2021arc} and spatial compatibility (ENI)~\cite{williams2022eni}.

\subsection{Input Space}

\begin{table*}[ht]
\centering
\resizebox{\textwidth}{!}{%
\begin{tabular}{|c|c|ccccccc|}
\hline
\rowcolor[HTML]{EFEFEF} 
\cellcolor[HTML]{EFEFEF} &
  \cellcolor[HTML]{EFEFEF} &
  \multicolumn{7}{c|}{\cellcolor[HTML]{EFEFEF}\textbf{Optimization Result}} \\ \cline{3-9} 
\rowcolor[HTML]{EFEFEF} 
\multirow{-2}{*}{\cellcolor[HTML]{EFEFEF}\textbf{\begin{tabular}[c]{@{}c@{}}Experiment\\ Conditions\end{tabular}}} &
  \multirow{-2}{*}{\cellcolor[HTML]{EFEFEF}\textbf{Sync Type}} &
  \multicolumn{1}{c|}{\cellcolor[HTML]{EFEFEF}\textbf{\begin{tabular}[c]{@{}c@{}}$\psi_{hor}$\end{tabular}}} &
  \multicolumn{1}{c|}{\cellcolor[HTML]{EFEFEF}\textbf{\begin{tabular}[c]{@{}c@{}}$\psi_{ver}$\end{tabular}}} &
  \multicolumn{1}{c|}{\cellcolor[HTML]{EFEFEF}\textbf{\begin{tabular}[c]{@{}c@{}}$\psi_{sem}$\end{tabular}}} &
  \multicolumn{1}{c|}{\cellcolor[HTML]{EFEFEF}\textbf{\begin{tabular}[c]{@{}c@{}}Registered \\ Space Size ($m^2$)\end{tabular}}} &
  \multicolumn{1}{c|}{\cellcolor[HTML]{EFEFEF}\textbf{\begin{tabular}[c]{@{}c@{}}Registered \\ Table Area Ratio\end{tabular}}} &
  \multicolumn{1}{c|}{\cellcolor[HTML]{EFEFEF}\textbf{\begin{tabular}[c]{@{}c@{}}Optimal RTGs\\ $(G = \left [ G_{x}, G_{y} \right ])$\end{tabular}}} &
  \textbf{\begin{tabular}[c]{@{}c@{}}Optimal \\ Position ($\phi = (x, y)$)\end{tabular}} \\ \hline
 &
  RTG-Grid &
  \multicolumn{1}{c|}{1} &
  \multicolumn{1}{c|}{0.2485} &
  \multicolumn{1}{c|}{0.8391} &
  \multicolumn{1}{c|}{27.04} &
  \multicolumn{1}{c|}{1} &
  \multicolumn{1}{c|}{\begin{tabular}[c]{@{}c@{}}{[}(1.0794, 1.0384, 1.0644),\\ (0.8860, 1.1178, 1.1260){]}\end{tabular}} &
  \begin{tabular}[c]{@{}c@{}}(0.7542, $-$0.1305)\end{tabular} \\ \cline{2-9}
 &
  RTG-Single &
  \multicolumn{1}{c|}{1} &
  \multicolumn{1}{c|}{0} &
  \multicolumn{1}{c|}{0.8201} &
  \multicolumn{1}{c|}{26.03} &
  \multicolumn{1}{c|}{1} &
  \multicolumn{1}{c|}{{[}1.0392, 1.0733{]}} &
  \begin{tabular}[c]{@{}c@{}}(0.7450, $-$0.1000)\end{tabular} \\ \cline{2-9} 
\multirow{-4}{*}{\begin{tabular}[c]{@{}c@{}}XR Studio\\ \& XR Lab \end{tabular}} &
  1:1 Scale &
  \multicolumn{1}{c|}{0.4796} &
  \multicolumn{1}{c|}{0} &
  \multicolumn{1}{c|}{0.8308} &
  \multicolumn{1}{c|}{24.14} &
  \multicolumn{1}{c|}{0.9032} &
  \multicolumn{1}{c|}{-} &
  \begin{tabular}[c]{@{}c@{}}(0.7500, $-$0.0500)\end{tabular} \\ \hline
 &
  RTG-Grid &
  \multicolumn{1}{c|}{0.4613} &
  \multicolumn{1}{c|}{0.1295} &
  \multicolumn{1}{c|}{0.5812} &
  \multicolumn{1}{c|}{31.33} &
  \multicolumn{1}{c|}{0.9076} &
  \multicolumn{1}{c|}{\begin{tabular}[c]{@{}c@{}}{[}(1.1297, 1.2510, 1.1188), \\ (1.0287, 1.1621, 1.1321){]}\end{tabular}} &
  \begin{tabular}[c]{@{}c@{}}(0.2375, $-$0.1231)\end{tabular} \\ \cline{2-9} 
 &
  RTG-Single &
  \multicolumn{1}{c|}{0.4612} &
  \multicolumn{1}{c|}{0} &
  \multicolumn{1}{c|}{0.5854} &
  \multicolumn{1}{c|}{24.81} &
  \multicolumn{1}{c|}{0.8914} &
  \multicolumn{1}{c|}{{[}1.2398, 1.0081{]}} &
  \begin{tabular}[c]{@{}c@{}}(0.1853, $-$0.2000)\end{tabular} \\ \cline{2-9} 
\multirow{-4}{*}{\begin{tabular}[c]{@{}c@{}}XR Studio\\ \& Meeting\\ Room\end{tabular}} &
  1:1 Scale &
  \multicolumn{1}{c|}{0.3875} &
  \multicolumn{1}{c|}{0} &
  \multicolumn{1}{c|}{0.6044} &
  \multicolumn{1}{c|}{25.86} &
  \multicolumn{1}{c|}{0.7298} &
  \multicolumn{1}{c|}{-} &
  \begin{tabular}[c]{@{}c@{}}(0.2521, $-$0.0979)\end{tabular} \\ \hline
 &
  RTG-Grid &
  \multicolumn{1}{c|}{0.4775} &
  \multicolumn{1}{c|}{0.1723} &
  \multicolumn{1}{c|}{0.5147} &
  \multicolumn{1}{c|}{38.26} &
  \multicolumn{1}{c|}{0.4731} &
  \multicolumn{1}{c|}{\begin{tabular}[c]{@{}c@{}}{[}(1.1260, 1.1436, 1.1343), \\ (0.9370, 0.9323, 1.1186){]}\end{tabular}} &
  \begin{tabular}[c]{@{}c@{}}(1.1081, $-$1.3088)\end{tabular} \\ \cline{2-9} 
 &
  RTG-Single &
  \multicolumn{1}{c|}{0.4759} &
  \multicolumn{1}{c|}{0.1565} &
  \multicolumn{1}{c|}{0.4890} &
  \multicolumn{1}{c|}{34.80} &
  \multicolumn{1}{c|}{0.4712} &
  \multicolumn{1}{c|}{{[}1.1365, 0.9481{]}} &
  \begin{tabular}[c]{@{}c@{}}(1.1131, $-$1.3024)\end{tabular} \\ \cline{2-9} 
\multirow{-4}{*}{\begin{tabular}[c]{@{}c@{}}XR Studio\\ \& Home\end{tabular}} &
  1:1 Scale &
  \multicolumn{1}{c|}{0.2551} &
  \multicolumn{1}{c|}{0} &
  \multicolumn{1}{c|}{0.4725} &
  \multicolumn{1}{c|}{31.93} &
  \multicolumn{1}{c|}{0.4146} &
  \multicolumn{1}{c|}{-} &
  \begin{tabular}[c]{@{}c@{}}(1.3486, $-$1.3046)\end{tabular} \\ \hline
 &
  RTG-Grid &
  \multicolumn{1}{c|}{0.5310} &
  \multicolumn{1}{c|}{0} &
  \multicolumn{1}{c|}{0.7921} &
  \multicolumn{1}{c|}{9.45} &
  \multicolumn{1}{c|}{0.5370} &
  \multicolumn{1}{c|}{\begin{tabular}[c]{@{}c@{}}{[}(1.0687, 1.1040, 1.0902), \\ (0.9017, 0.8995, 0.9283){]}\end{tabular}} &
  \begin{tabular}[c]{@{}c@{}}(0.7516, 0.8928)\end{tabular} \\ \cline{2-9} 
 &
  RTG-Single &
  \multicolumn{1}{c|}{0.5333} &
  \multicolumn{1}{c|}{0} &
  \multicolumn{1}{c|}{0.7924} &
  \multicolumn{1}{c|}{9.41} &
  \multicolumn{1}{c|}{0.5398} &
  \multicolumn{1}{c|}{{[}1.1067, 0.8997{]}} &
  \begin{tabular}[c]{@{}c@{}}(0.7510, 0.8952)\end{tabular} \\ \cline{2-9} 
\multirow{-4}{*}{\begin{tabular}[c]{@{}c@{}}XR Studio\\ \& Office\end{tabular}} &
  1:1 Scale &
  \multicolumn{1}{c|}{0.2857} &
  \multicolumn{1}{c|}{0} &
  \multicolumn{1}{c|}{0.7746} &
  \multicolumn{1}{c|}{9.45} &
  \multicolumn{1}{c|}{0.4878} &
  \multicolumn{1}{c|}{-} &
  \begin{tabular}[c]{@{}c@{}}(0.7500, 0.9000)\end{tabular} \\ \hline
\end{tabular}
}
\caption{The table shows the optimization results of three registration methods for four evaluation conditions: horizontal edge match ratio about the table ($\psi_{hor}$), vertical plane match ratio about walls ($\psi_{ver}$), semantic match ratio ($\psi_{sem}$), registered space size and registered table area ratio. Optimal RTGs (G) and optimal position ($\phi$) for rescaling and registering physical space to the XR Studio are also written.}
\label{tab:result}
\end{table*}

Our evaluation procedure involved generating input 3D spaces for AR/VR remote collaboration scenarios, where VR clients from different remote locations could collaborate with AR hosts in an XR Studio. The XR Studio, which was equipped with three large monitors, tables, PC desks, and a meeting table with chairs, was set as a target virtual space. \autoref{fig:dataset} shows the 3D model of the XR Studio, which is similar to telepresence components in the Cisco TelePresence IX5000 Series\footnote{https://www.cisco.com/c/en/us/products/collaboration-endpoints/immersive-telePresence}. Our XR meeting scenario designated the table in front of the three monitors as the main object. \autoref{fig:dataset} shows extracted floorplans from the top view of input spaces and preprocessed to polygon based on the bounding box with three semantic labels. The first label, ``floor," indicates where users can walk in the space. The second label, ``main object," identifies the object used as the center of virtual content. The third label, ``obstacle," was assigned to the remaining objects in physical and virtual spaces, as users cannot move around areas where objects exist. \autoref{fig:dataset} shows each label is drawn with different boundary colors where the floor area is black, the main object is orange, and the obstacles are red. 

\autoref{fig:dataset} shows the top view of four selected physical spaces' 3D models from real spaces labeled with the same regulation as the XR Studio. The XR Lab has a similar layout to XR Studio but with differences in the number of obstacles and space size. The Meeting Room had a wide table in the middle and chairs around it, and another small desk and chairs were placed in the corner. The Home was made from a general living room with a sofa, and the area of the main object was approximately half of that in XR Studio. Lastly, the Office was made from a single-person workplace that has a different spatial configuration from XR Studio in every aspect. We set the largest table from each physical space as the main object and assumed each table has the same height to focus on generating the horizontal redirection map with RDW.

\subsection{Experiment Design}

\textbf{Spatial Registration}: In order to evaluate the effectiveness of our proposed space registration method (RTG-Grid), we compared RTG-Grid with two other methods: the 1:1 Scale method, which matches spaces at a one-to-one scale without modifying it~\cite{keshavarzi2020optimization}, and the Single-RTG method, which applies a single RTG to the entire physical space~\cite{kim2022mutual}. To evaluate the three space registration methods' performance, we generated an objective function with \autoref{equ:objFunc}, which has a single $\psi_{hor}$ about the table and a single $\psi_{ver}$ about the room's walls. We multiplied the host's space area to the $\psi_{size}$ for a more intuitive result representation and displayed the registered space size. Moreover, the registered table area ratio was also computed to compare the utilizable table surface between the methods. Finally, the weights for two edge-centric evaluation metrics ($\psi_{hor}$, $\psi_{ver}$) were higher than those of $\psi_{size}$ and $\psi_{sem}$. The same objective function was used for three spatial registration methods, and we will validate the performance of the RTG-Grid by comparing the evaluation metrics' values and visually intersected results.

\begin{table}[]
\centering
\resizebox{\columnwidth}{!}{%
\begin{tabular}{@{}cccccc@{}}
\toprule
Space        & Size ($m^2$) & Table Size ($m^2$) & C($E$) & OS($E$) & SC($E$) \\ \midrule
XR Studio    & 8.4 $\times$ 6.0 & 4.10 $\times$ 0.80  & 6.50 & 69.18 & 115.30\\
XR Lab       & 6.8 $\times$ 3.6 & 3.95 $\times$ 0.75 & 7.36 & 6.68  & 43.04\\
Meeting Room & 7.6 $\times$ 3.5        & 3.00 $\times$ 1.20 & 6.64 & 14.56 & 48.80\\
Home         & 6.8 $\times$ 6.0        & 1.70 $\times$ 0.85 & 8.55 & 7.27  & 57.92\\
Office       & 3.5 $\times$ 2.7        & 2.00 $\times$ 0.90 & 3.30 & 9.27 & 17.42\\ \bottomrule
\end{tabular}%
}
\caption{Five input spaces' size, table (main object) surface size, the complexity of an environment C($E$)~\cite{williams2021arc}, object scatteredness OS($E$), and spatial complexity SC($E$).}
\label{tab:space}
\end{table}

\textbf{Spatial Dissimilarity}: We compared proposed SD and SMD with Complexity Ratio (CR) ~\cite{williams2021arc} and Environmental Navigation Incompatibility (ENI) ~\cite{williams2022eni} about given space sets to understand their characteristics in measuring spatial complexity. \autoref{tab:space} displays the size of each space and the main object's surface for the five input spaces. We calculated C($E$) using a point sampling with every 0.2 m~\cite{williams2021arc} and obtained OS($E$) with \autoref{equ:OS}, which are also listed in \autoref{tab:space}. Finally, the spatial complexity for each space can be obtained with \autoref{equ:SC} and also written in \autoref{tab:space}. Since ENI is highly affected by the physical space's complexity, we added one more physical space named Simple, which has a 10 m $\times$ 10 m size with a 2 m $\times$ 2 m table in the center of the space to validate the difference between ENI and SD clearly.

\subsection{System Implementation}

Our evaluations were conducted on a MacBook Pro 13-inch Monterey OS with Apple M1, 8-core CPU, 16GB RAM, and 512GB SSD. We implemented our algorithm with the python optimization library Pygmo 2.19.0~\cite{Biscani2020} in the Anaconda virtual environment~\cite{anaconda} of python 3.9.13 and utilized a Pypex\footnote{https://github.com/mikecokina/pypex/} for work with polygons in 2D. As RTG-Grid has four inequality constraints for six input variables (six translation gains) and RTG-Single has two inequality constraints for two input variables (two translation gains), we used Pygmo's meta-optimization algorithm based on Farmani et al.~\cite{farmani2003self}'s self-adaptive fitness formulation. For a 1:1 scale method, the extended ant colony optimization algorithm~\cite{dorigo2019ant} was used since the input variable for the 1:1 scale's input variables (position vector) has no inequality constraints between them. To visualize the spatial matching results, we used Python's Pillow (PIL) 9.4.0 library~\cite{umesh2012image}. The edge match function determined as the edge was matched when the identical labeled polygons' edges from the physical and virtual space were parallel and less than 0.01 m since a few centimeters can be precisely matched by using the hand redirection techniques~\cite{zenner2019estimating, gonzalez2022model}.

The input data for our spatial matching optimization algorithm is shown in \autoref{fig:dataset}'s polygonal input data row. Each polygonal input data were represented as a list of points. To identify whether polygons are intersected and the area of an intersected polygon, the polygon-clipping method was used~\cite{vatti1992generic}. Since our evaluation assumed table-centric XR collaboration, the weight of the objective function \autoref{equ:objFunc} was set as $\omega_{1,1}$ = 100, $\omega_{2,1}$ = 30, $\omega_{3}$ = 5, and $\omega_{4}$ = 10 to match the edge and plane primarily. Since XR Studio has a similar configuration with Medium $\times$ Furnished from Kim et al.~\cite{kim2022configuration}'s study, we set the RTG threshold $\alpha_{T, l}$ as 0.92 and $\alpha_{T, h}$ as 1.23. For redirection gain smoothing, we set $l_s$ as 0.25 m.

\subsection{Result}

\begin{figure*}[ht]
 \centering
 \includegraphics[width=\textwidth]{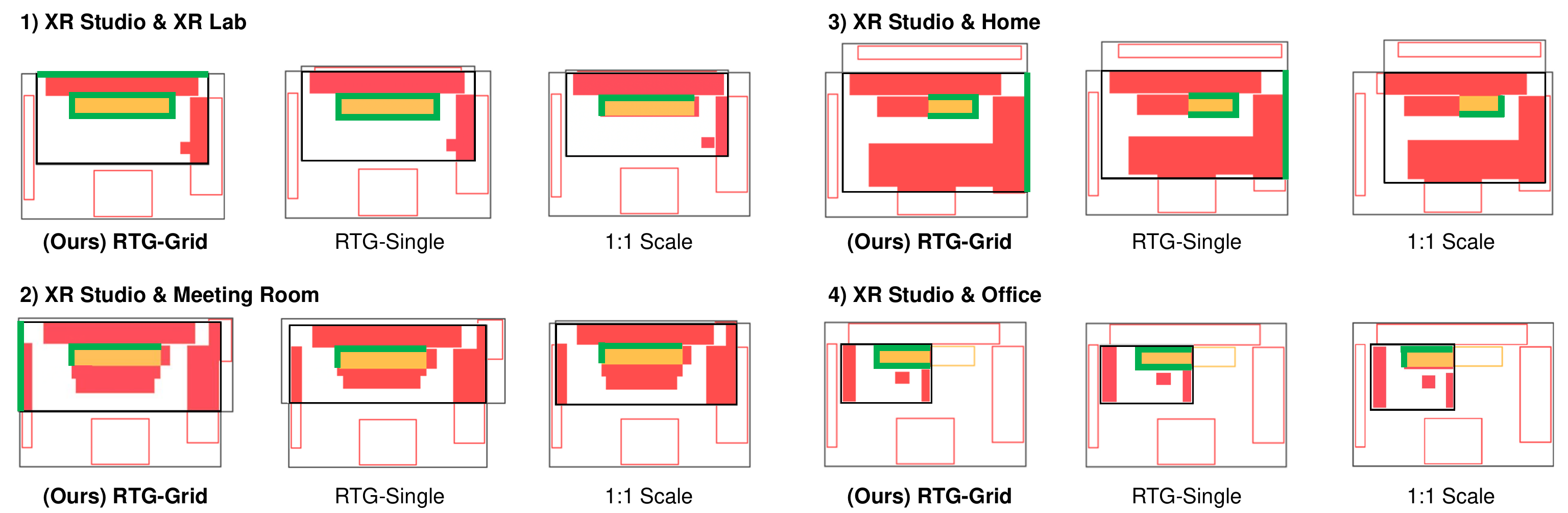}
 \caption{Space registering results of four input space pairs. The transparent polygons are an intersection of two spaces, and the black rectangle overlayed on it refers to the registered area between physical and virtual spaces. The green lines indicate the aligned table edge and the wall surface, the orange areas indicate the registered table surface, and the red areas indicate the registered obstacles or label-unmatched areas.}
 \label{fig:result}
\end{figure*}

\textbf{Spatial Registration}: \autoref{tab:result} shows the optimization results for four pairs of physical and virtual spaces. \autoref{fig:result} shows the space registering results with optimal values where the transparent polygons refer to an intersected space. The green lines indicate the aligned table edge and the wall surface, the orange areas indicate the registered table surface, and the red areas indicate the registered obstacles or label-unmatched areas. For the first pair, XR Lab and XR Studio, \autoref{tab:result} shows that every table edge of the virtual main table was matched to the physical main table with RTG-Grid and RTG-Single, which is more than double compared to the 1:1 Scale method. With RTG-Grid and RTG-Single, all surfaces of the virtual table were interactable, and 90.32\% of them were usable in the 1:1 Scale. Additionally, only RTG-Grid could align 24.85\% of the physical and virtual wall on the top side, as shown in \autoref{fig:result}. RTG-Grid had the highest semantic match ratio at 83.91\%, followed by 1:1 Scale at 83.08\%, and RTG-Single at 82.01\%. Finally, the registered space size was the largest with RTG-Grid at $27.04\ m^2$, followed by $26.03\ m^2$ in RTG-Single and $24.14\ m^2$ in 1:1 Scale.

The evaluation of the second space pair, consisting of the XR Studio and Meeting Room, is presented in \autoref{tab:result}. The table edge match ratio for the RTG-Grid was 46.13\%, while for RTG-Single and 1:1 Scale, it was 46.12\% and 38.75\%, respectively. Furthermore, only RTG-Grid could align 12.95\% of the physical and virtual wall on the left side, as shown in \autoref{fig:result}'s intersection result. For the registered table area ratio, RTG-Grid and RTG-Single align 90.76\% and 89.14\%, respectively, while 72.98\% of the table surface was usable in the 1:1 Scale. The semantic match ratio was highest in the 1:1 Scale at 60.44\%, followed by RTG-Single at 58.54\%, and RTG-Grid at 58.12\%. Finally, the size of the registered space was largest in RTG-Grid at $31.33\ m^2$, followed by $25.86\ m^2$ in 1:1 Scale and $24.81\ m^2$ in RTG-Single.

For the third space pair, XR Studio and Home, \autoref{tab:result} shows that the table edge match ratio of RTG-Grid was 47.75\%, while that of RTG-Single was 47.59\% and 1:1 Scale was 25.51\%. Using RTG, it was possible to align the table edge over 80\% more than that of the 1:1 Scale. Furthermore, \autoref{fig:result}'s $\psi_{ver}$ column showed that only RTG-Grid and RTG-Single could align 17.23\% and 15.65\% of the physical wall, respectively. The registered table area ratio was about 47\% with RTG-Grid and RTG-Single methods, while it was 41.46\% with the 1:1 Scale. Regarding the semantic match ratio, RTG-Grid was the highest at 51.47\%, followed by RTG-Single at 48.90\%, and 1:1 Scale at 47.25\%. Finally, the registered space size was the largest in RTG-Grid at $38.26\ m^2$, followed by $34.80\ m^2$ in RTG-Single and $31.93\ m^2$ in 1:1 Scale.

In the XR Studio \& Office space pair, \autoref{tab:result} shows the table edge match ratio for RTG-Grid and RTG-Single was about 53\%. In contrast, the 1:1 Scale achieved only 28.57\%. RTG methods achieved about an 85\% higher match ratio in aligning the edge between physical and virtual tables than the 1:1 Scale. None of the space registration methods could align the wall between the two spaces. The RTG-Grid and RTG-Single methods could align about 54\% of the table surface, while the 1:1 Scale allowed 48.78\%. Furthermore, the semantic match ratio was highest in RTG-Grid and RTG-Single, with about 79\% semantic match ratio achieved, while the 1:1 Scale reached 77.46\%. Finally, the registered space size of the XR Studio \& Office space pair was the smallest among the four space pairs, with RTG-Grid and 1:1 Scale having the registered space size at $9.45\ m^2$, followed by $9.41\ m^2$ in RTG-Single. 

\begin{table}[]
\centering
\resizebox{\columnwidth}{!}{%
\begin{tabular}{@{}cccccc@{}}
\toprule
Space Pairs               & CR & ENI & SD & SMD  \\ \midrule
XR Studio \& XR Lab       & 1.1323                   & 6.1989 & 0.9841                   & 0.0985 \\
XR Studio \& Meeting Room & 1.0215                   & 8.4415 & 0.8598                   & 0.2442 \\
XR Studio \& Home         & 1.3154                   & 3.7831 & 0.6884                   & 0.4464 \\
XR Studio \& Office       & 0.5077                   & 12.6301 & 1.8900                   & 1.2367 \\ 
XR Studio \& Simple       & 2.7266                   & 0.0055 & 3.3056                   & 0.9972 \\ \bottomrule
\end{tabular}%
}
\caption{CR~\cite{williams2021arc}, ENI~\cite{williams2022eni}, SD, and SMD for given space pairs}
\label{tab:com}
\end{table}

\textbf{Spatial Dissimilarity}: The CR~\cite{williams2021arc}, ENI~\cite{williams2022eni}, SD, and SMD for each space pair were shown in \autoref{tab:com}. CR indicates two spaces are more similar when it is close to 1, and ENI refers to the spatial compatibility is higher when the value is closer to 0. SD and SMD mean spaces are similar and easy to register when close to 0. The first space pair, XR Studio \& XR Lab, resulted in a CR of 1.1323, ENI of 6.1989, SD of 0.9841, and SMD of 0.0985. For the second pair, XR Studio \& Meeting Room, CR was 1.0215, ENI was 8.4415, SD was 0.8598, and SMD was 0.2442. In the third pair, XR Studio \& Home, CR was 1.3154, ENI was 3.7831, SD was 0.6884, and SMD was 0.4464. In the fourth pair, XR Studio \& Office, CR was 0.5077, ENI was 12.6301, SD was 1.8900, and SMD was 1.2367. Finally, in the last additional pair, XR Studio \& Simple, CR was 2.7266, ENI was 0.0055, SD was 3.3056, and SMD was 0.9972.

\subsection{Discussion}

\autoref{tab:result} demonstrates that using RTG to rescale spatial dimensions improved the accuracy of matching table edges and walls in all four experimental conditions compared to the 1:1 Scale method. RTG-Single and RTG-Grid had comparable performance in matching table edges, but RTG-Grid outperformed RTG-Single in matching both table edges and walls. In addition, RTG-Grid yielded the largest registered space area in all experimental conditions, signifying that users had more space to interact with. By comparing RTG-Grid's results with other methods, we verified that our approach accurately registered the virtual space to the physical space while ensuring users' unrestricted mobility. Furthermore, all three spatial registration methods show the best $\psi_{sem}$ in at least one space pair, indicating that relying solely on the semantic match ratio is insufficient for evaluating spatial registration outcomes.

\autoref{tab:com} shows a comparison of proposed SD and SMD with spatial compatibility metrics CR~\cite{williams2021arc} and ENI~\cite{williams2022eni}. Since it is difficult to define the ground truth of the inter-space heterogeneity that takes into account all the elements of a space, we found that SD and SMD better reflect the spatial complexity characteristics of the given spaces compared to previous methods. According to \autoref{tab:space}, the C($E$) for XR Studio and Meeting Room was almost the same at 6.5 and 6.64, respectively. Although these two spaces had different spatial configurations, the CR suggests that XR Studio and Meeting Room have similar spatial complexity. On the other hand, the SC($E$) from \autoref{tab:space} shows the difference between XR Studio and the Meeting Room where the SC($E$) of XR Studio was 115.30, and that of the Meeting Room was 48.80. These results support that SC($E$) could capture similarity in terms of spatial configuration better than C($E$). Additionally, when we arranged the space conditions in ascending order of SC($E$), Office was found to be the most complex, followed by XR Lab, Meeting Room, Home, and XR Studio. This result was consistent with previous research that qualitatively ordered various space conditions based on spatial complexity~\cite{shin2022effects}. Thus, we confirmed that SC($E$) could reflect spatial dissimilarity in terms of object distribution better than C($E$). On the other hand, \autoref{tab:com} shows the ENI score for XR Studio \& Simple was 0.0055, even though the two spaces are highly dissimilar. This is because ENI measures how easy it is to move around in a virtual space from the user's physical space. However, SD measures the spatial difference in terms of spatial configuration, so \autoref{tab:com} shows SD and SMD to XR Studio \& Simple show two spaces are highly dissimilar. Therefore, the ENI is appropriate when applying RDW to navigate the entire VR space, but to measure spatial dissimilarity considering spatial layout and object density, our proposed SD and SMD are more suitable.

Spatial registration results show that the spatial registration results tended to improve when SMD decreases, indicating that the SMD is an appropriate measure to evaluate the difficulty of object-centric physical-virtual space registration. \autoref{tab:com} shows that the SD of XR Studio \& Home is lower than the XR Studio \& Meeting Room, but the SMD of XR Studio \& Home is higher than the XR Studio \& Meeting Room. \autoref{tab:com} also reveals that the lowest SMD was observed for XR Studio \& XR Lab, followed by XR Studio \& Meeting Room, XR Studio \& Home, and XR Studio \& Office, implying that the spatial matching difficulty increased in this order. In the case of XR Studio \& XR Lab, RTG-Grid and RTG-Single successfully aligned the horizontal-/vertical plane, and also the semantic match ratio and registered space area were highest compared to other conditions. XR Studio \& Meeting Room followed the next SMD level, and it shows 46\% of the table edge match ratio and more than 90\% of the virtual table surface was registered. The XR Studio \& Home has the third SMD level, in which the table edge can be matched about 48\% with RTG-Grid, and it's similar to that of XR Studio \& Meeting Room. However, the registered table area is only 47\%, which is about half the size of the registered table area from XR Studio \& Meeting Room. Finally, XR Studio \& Office had the highest SMD value of 1.2367, leading to almost one-third of the registered space area compared to other evaluation conditions. Although space-matching results were hard to define as a total score, the tendency of spatial matching results decreases according to the SMD value increases indicating that the SMD is a suitable measure to evaluate the difficulty of aligning physical and virtual spaces.

Based on the findings, we could derive three implications for developers implementing optimization algorithms for registering virtual environments to a user's physical space. The first implication is to use RDW as a space-adaptive rescaling method for virtual-physical space registration. Our results confirmed that RTG could improve the matching ratio of edge and plane information between the two spaces. Therefore, utilizing the RTG in RDW for accurately registering virtual and physical spaces is recommended. Next, the primary component of space registration should be determined considering the interaction with the virtual main object. When accessing a space, it can be divided into two main parts: perceivable and interactable areas. When physical and virtual spaces are highly dissimilar, it is more important to precisely match the interactable objects rather than inaccurately match the overall space. The last implication is focusing on registering a specific subspace of virtual spaces to the physical space. This involves designating a proper subspace as the interactable space to register the virtual environment to the physical space. This divide-and-conquer approach can overcome the spatial dissimilarity between physical and virtual spaces and also have the potential to be used for multi-space allocation for further studies.

\section{Conclusion and Future Work}

This study proposed a novel edge-centric physical space rescaling technique with RDW to register dissimilar virtual-physical spaces. We developed spatial matching metrics to measure the edge and plane match ratio between physical and virtual spaces and presented four quantitative evaluation metrics for spatial registration. Our optimization algorithm with \autoref{equ:objFunc} determines the optimal RTGs and positions according to the main interactable object utilized mainly in the target VR scenario. Furthermore, we introduced new spatial metrics, spatial complexity SC($E$), spatial dissimilarity SD($E_{virt}, E_{phys}$), and spatial matching difficulty SMD($E_{virt}, E_{phys}$), to help design more structured studies and validate corresponding results. Furthermore, our evaluation with previous methods (RTG-Single, 1:1 Scale) demonstrated that our method (RTG-Grid) was beneficial in registering more edge, plane, and movable area compared to other spatial matching methods. Lastly, considering the target scenario, we derived three implications for developers to consider when designing a spatial matching algorithm for registering dissimilar virtual spaces to real spaces.

To the best of our knowledge, our research is novel in proposing using RDW for adaptive space rescaling to register virtual space to physical space precisely. However, some limitations exist. Firstly, a user study is necessary to verify the effectiveness of our spatial registration results, as we anticipate that supporting physical touch in VR by aligning the edge and plane of physical space would improve user immersion. Additionally, the effects of gradual gain changes on user experience should be explored. Next, we should apply our 2D horizontal retargeting map in conjunction with the 3D hand redirection technology. In our experiment, we assumed the height of the table is the same to focus on horizontal redirection, but horizontal surfaces' heights will be different in many cases. In order to handle this height difference between objects, we should apply the hand redirection technique by considering the position of users together. Finally, this study is also limited by the fact that the space is divided around a single target horizontal surface from the table and only considers four specific room pairs.

In a follow-up study, we will first conduct a pilot study on the effectiveness of smooth gain change and physical interaction on a single object, such as a desk. We will design more specific user experiments based on the pilot study results. Secondly, a 3D spatial redirection should be developed by integrating the hand redirection technique into the horizontal redirection map generated with our method. This could generate a more reliable and accurate redirection map that could handle cases where the height of the interactable plane in the physical space differs from that of the virtual space. Finally, extending the algorithm to adaptively partition and deform the space when there are multiple interacting objects is necessary. The algorithm's performance needs to be validated on a wider variety of spatial pairs. With the above features and validation, we can create a universal-level algorithm contributing to immersive VR space experiences.

\acknowledgments{
This research was supported by National Research Council of Science and Technology (NST) funded by the Ministry of Science and ICT (MSIT), Republic of Korea (Grant No. CRC 21011).
}


\bibliographystyle{abbrv-doi-hyperref}

\bibliography{template}
\end{document}